\documentclass[12pt]{iopart}
\usepackage{iopams}

\newcommand{\be}{\begin{equation}}
\newcommand{\ee}{\end{equation}}
\newcommand{\ba}{\begin{eqnarray}}
\newcommand{\ea}{\end{eqnarray}}
\newcommand{\n}{\nonumber}
\newcommand{\lan}{\langle}
\newcommand{\ran}{\rangle}
\newcommand{\B}{\beta}
\newcommand{\A}{\alpha}
\newcommand{\w}{\gamma}
\newcommand{\s}{\sigma}
\newcommand{\z}{\zeta}
\newcommand{\ww}{\widetilde{\gamma}}
\newcommand{\defeq}{\mathrel{\mathop:}=}
\newcommand{\eqdef}{=\mathrel{\mathop:}}

\begin{document}

\title{Reposition time in probabilistic imperfect memories}
 \author{Ludmi{\l}a Praxmeyer}
\address{Department of Physics,  Sofia University, James Bourchier 5 blvd, 1164 Sofia, Bulgaria}
\ead{lpraxm@gmail.com}

\begin{abstract}
We present the analysis of behavior of $N$ identical finite time 
memories with the imperfections characterized by a step-function 
model, based on a study of independent copies of the geometric 
distribution. We show a step-by-step derivation of a formula for the 
average waiting time for obtaining $N$ successes  within a given 
time span~$\tau$, which was used recently in the analysis of quantum 
repeater rates. 
\end{abstract}


\section{Introduction}

The problem we are solving in this paper is motivated by a 
stochastic behavior of physical objects studied in the context of 
quantum information processing or, more precisely, in the quantum 
repeater rates analysis. Although in next paragraphs  a general 
description of the system that inspired questions posed and answered 
here is given, familiarity with quantum information theory is not 
needed to follow the main part of this paper. Presented results  are 
completely independent of the physical context and  were obtained 
employing minimal mathematical tools: geometric series, multinomial  
coefficients and elementary probability theory (i.e.,  knowledge on 
the level of two first chapters of~\cite{Feller}). A slight 
complication comes from the fact that subsequently defined 
probability spaces are constructed from ones introduced previously. 
As a consequence, similar but connected with different spaces  terms 
and expressions  appear in the text. To avoid confusion, we  employ 
rather excessive notation in order to always keep track of random 
variables and sample spaces. Basic information on the notation can 
be found  in {\mbox{\ref{A_A1}}}. Throughout the   paper term `game' 
is used to describe sequences of motions guided by earlier defined 
rules, which is not the same meaning as the word usually has in the 
sense of game theory. Readers interested mainly in  mathematical 
aspects of this work might want to skip the rest of this 
introduction and go directly to section 2.

When dealing with quantum systems, a large part of experimental
challenges  comes from the relation between the time scale of
coherence stability and the time needed for  effective, practically
useful manipulations on the system. Difficulty with maintaining
coherence is directly connected  with the   problem  of quality of 
quantum memories {\mbox{-- and approaching}} the subject from this 
angle one can apply many classical intuitions. On the most 
elementary level, memory can be treated as a  function of time with 
only two possible values:
 `success', when the state is faithfully stored, or
`failure' corresponding to the unsuccessful storage attempt. In this
  paper, we use such a simple step-function model assuming that memory
is perfect for a specific time $\tau$, but after that time the stored
state is completely and irreversibly lost and memory returns to its
initial state. Although in the description of noise and decoherence
exponential decay is more common,  it can always be reduced to the
step-function model by adding a threshold and a cut-off,  to mimic
the assumption that a perfect to some point memory  rapidly
 loses the ability to store the state.

As mentioned before, the initial motivation for this study was the 
need to develop analytical tools for the rates analysis of quantum 
repeaters, i.e., devices designed to distribute entanglement over 
large distances in a more efficient way then the direct (quantum) 
communication between two  remote locations. The main idea of 
quantum repeater is  to divide initial distance into many segments, 
establish entanglement over these shorter distances,  then connect 
them by entanglement swapping, usually  adding also steps of 
entanglement purification  \cite{qrep}.
 Depending on a
repeater model number of segments, or nodes per segments, etc.
changes; depending on a chosen strategy  procedures of purification
and  swapping might appear in different order.  Nevertheless,
  in every repeater model, no mater of its construction
strategy,  a successful creation of the initial `small-distance'
entanglement is always a starting point
 and a necessary condition for
establishing the long-distance entanglement.  Discussion in this 
paper is restricted to this initial process and focuses on effects 
of memory imperfections. 

We will consider the following problem: Imagine, that a single 
attempt of entanglement creation  in one segment of quantum repeater 
ends successfully with probability $p$. Once created, entanglement
 is stored perfectly for a specific time $\tau$ and
after that time it gets irreversibly lost. We are interested in 
finding the average waiting times for creation of  entanglement  in  
$N$ identical and independent segments. To be on a save side and 
rather underestimate the rates then overestimate them, we assume 
that if entanglement was lost in one segment before it was created 
in all the others,  the whole system is reset to its initial state. 
If this happens, a  `residual' entanglement stored in any segment is 
discarded, and attempt of creating it in all of the segments starts 
from the beginning. Defined in this way, it is a purely 
probabilistic  problem that has an exact analytical solution. For 
the perfect memories (infinite $\tau$) the solution was shown in
\cite{Nadja1}; for finite memory times $\tau$ we present it in this paper.

We have put some effort to obtain analytical solutions in order to 
be able to establish bounds, calculate accurate limits, and easily 
compare different systems. Comparison between two analytical 
formulas is  unimpeachable,  whereas the same can not be said 
unconditionally about numerical results. Analytical solutions build 
intuitions and  understanding of the system behavior; they allow for 
formal proofs -- in contrast to educated guesses. As stated before, 
results presented here are independent from physical settings.  They 
were developed in the context of repeater analysis
\cite{Nadja2}; but describe, as well, any other 
$N$-components system with the characteristic as given above. 
Formulas obtained are universal and their simple structure makes 
them convenient for implementations.

The   paper is structured as follows: In section 2, general rules of 
the games considered here are defined;  a detailed description of 
the system studied and the problem we are solving is given. 
Sections~3--5 are devoted to the systematic presentation of the 
solution of this problem. Most of the details concerning 
calculations and notation  are moved to the Appendixes.

\section{System description}
Consider a classical bit with values (states) denoted by $|0\ran$ 
and $|1\ran$. Let  $T$ be a process that with a probability~$p$ $ 
(0<p<1) $ transfers a bit from a state $|0\ran$ to $|1\ran$, but 
acting on a bit in a state $|1\ran$ always leaves it untouched. A 
pair constructed from a sample space $ 
{\mathbb{A}}=\{s_{\mathbb{A}},f_{\!\mathbb{A}}\} $ and a map $ 
\A:2^{\{s_{\mathbb{A}},f_{\mathbb{A}}\}}
\longrightarrow [0,1]\,  $, characterized by the properties:
\ba 
\A(\{\emptyset\}):=0  \qquad
\A(\{s_{\mathbb{A}},f_{\!\mathbb{A}}\}):=1
\qquad\A(s_{\mathbb{A}}):=p\qquad \A(f_{\!\mathbb{A}}):=1-p=q,
\n
\ea
defines the probability space, which describes formally  action of 
$T$ on $|0\ran$. The event $s_{\mathbb{A}}$ occurs with the 
probability $p$, and corresponds to the successful bit-flip from 
$|0\ran$  to~$|1\ran$. The event $f_{\mathbb{A}}$ happens with the 
probability {\mbox{$q=1-p$}}, and corresponds to the failed attempt 
of a state transfer.

In this paper,  we  consider systems composed of $N$ classical bits,
all initiated in states~$|0\ran$, and probabilistic `games' during
which one tries to transfer these bits into states~$|1\ran$,
repeatedly applying the procedure $T$. The action of $T$ on the
$N$-bit
 system is defined as the simultaneous and independent action of
$T$ on every bit:
\ba
 T(bit_1,bit_2, ... , bit_N) \defeq
(T(bit_1),T(bit_2),... ,T(bit_N))\,. \n
\ea

\subsection{Game ${\cal{G}}_N$, perfect memories }\label{section2.1}

Let ${\cal{G}}_N$ be a game, in which the goal is to transfer  $N$
bits from states $|0\ran$ into $|1\ran$ by repeating applications of
the procedure $T$. A question of an average number of repetition 
needed to obtain this goal was answered in \cite{Nadja1} 
(rediscovering result from
\cite{Szpank} and \cite{Eisenberg}). For fixed
$N$ and $q=1-p\,$, it is expressed by
 \ba
 \lan \mathcal{K}_{{\cal{G}}_N} \ran = \sum_{k=1}^{N}
 { \displaystyle{{N \choose k}}}\frac{(-1)^{k+1}}{1-q^k} \label{perfect}
\,.
\ea
This result can be generalized: solving recursive formulas, obtained
similarly as in \cite{Nadja1}, we find that an average number of
applications  of $T$  needed to transfer at least $m$ from $N$ bits
of the system from $|0\ran$ into $|1\ran$
 is given by
\ba
\lan \mathcal{K}_{m,N}\ran=\frac{1}{(m-1)!} \sum_{k=0}^{m-1} \frac{(-1)^{m-k}}{(m-k)}
{m-1
\choose k}
\frac{N!}{(N-m)!} \frac{q^k}{q^N-q^k}\,.\n 
\ea
It is worth noting, that for $0<p< 1$, after sufficiently many
applications of the procedure $T$, all $N$ bits are transferred to
the state $|1\ran$. Because of the implied in the definition of
${\cal{G}}_N$  (and quite reasonable for classical systems)
assumption that the bits in state~$|1\ran $ stay
 undisturbed, no matter how many repetitions of $T$
followed --  we say it is a perfect memory case.

\subsection{Games  ${\cal{G}}_{(N,\tau)}$, imperfect memories  
(finite memory time $\tau$) }\label{section2.2}

A more physical (at least for quantum systems) model should take
into account possibility that bits in state $|1\ran$  are left
undisturbed only for some finite time $\tau$.   For simplicity, we
will use here discrete time -- measured in steps -- with the number
of steps equal to the number of the subsequent applications of
procedure $T$. A memory time $\tau$ means that the system stability
can be guaranteed only for $\tau$ repetitions of $T$, counting from
the first successful transfer  to  state~$|1\ran$.  After that time
the bit (or bits), which flipped to~$|1\ran$ as first, is no longer
in a well defined state. This is the reason why only those events,
when all $N$ bits flipped to~$|1\ran$ within the time span~$\tau$,
will be considered as successes in the modified version of
game~${\cal{G}}_N$.

Obviously, an addition of the assumption of imperfect memories to
the game ${\cal{G}}_N$ changes its rules and results. Let us fix
parameters $N$ and $\tau$. A modified game ${\cal{G}}_{(N,\tau)}$ is
divided in rounds, which always start with $N$ bits prepared in
states $|0\ran$. During a single round, procedure $T$ is applied to
the whole system until all bits  are transferred to states $|1\ran$
but not more then $\tau$ times, counting from the first successful
transfer of at least one bit  into state $|1\ran$.  If time~$\tau$
was not enough to transfer all bits into states~$|1\ran$, the round
ends unsuccessfully; the system is reset to its initial
state~$|0\ran$ and  the next round starts, automatically. The game
 ends when all bits are finally
transferred to $|1\ran$, i.e.,  in a moment when some round ended
successfully. Note that, for $\tau \geq 1$,  game
${\cal{G}}_{(N,\tau)}$ does not form a Markov chain. We would like
to know, how many times on average procedure~$T$ is  repeated  in
game ${\cal{G}}_{(N,\tau)}$. Answer,  given by  the formula
\ba
\lan \mathcal{K}_{{\cal{G}}_{(N,\tau)}} \ran= \frac{ 1-(1-q^\tau)^N +
(1-q^N)\Big[\tau-\sum_{j=1}^{\tau-1}(1-q^j)^N \Big]
}{(1-q^{\tau+1})^N-q^N(1-q^\tau)^N}\,, \n 
\ea
and its derivation are the main results of this paper. 

When considering a single round of game ${\cal{G}}_{(N,\tau)}$, it is
convenient to divide it into two separate parts. Initial $k\geq 0$
steps ($k$ applications of procedure~$T$), during which all $N$ bits
remain in state $|0\ran$, constitute the first part.  The second
part starts at step $k+1$, when at least one bit gets transferred to
$|1\ran$. A course of the second part is completely independent from
a duration of the first. A general description of the second part of
an arbitrary round is presented in  section
\ref{section3}.
 In section
\ref{section4},  both of these parts are combined and the general
description of single rounds of ${\cal{G}}_{(N,\tau)}$ is
addressed.  Section
\ref{section5} is devoted to the results that characterize the whole game
${\cal{G}}_{(N,\tau)}$.

\section{Auxiliary spaces $ \Gamma(N,\tau) $}\label{section3}

In this section, we shall focus on the description of a second part
of an arbitrary round of  ${\cal{G}}_{(N,\tau)}$, i.e., the part
that starts with the first successful bit-flip. To keep track of all
possible events, we introduce auxiliary spaces $\Gamma(N,\tau) $
 defining them, for natural numbers  $\tau\geq 0$ and $N\geq 2$, by
\ba {\Gamma}(N,\tau)\defeq \Big\{{\gamma}\in
{\mathrm{Map}}((\overline{0,\tau}),(\overline{0,N})): {\gamma}(0)> 
0, \sum_{j=0}^\tau {\gamma}(j)\leq N \Big\}. 
\n
\ea
A map ${\gamma}\in{\Gamma}(N,\tau)$ corresponds to these rounds of 
${\cal{G}}_{(N,\tau)}$ during which  all $N$ bits stayed in state 
$|0\ran$ for initial ${k}$ steps, then at step ${k}+1$ at least one 
bit (${\gamma}(0)>0$ bits) flipped from state $|0\ran$ to $|1\ran$,  
at step ${k}+2$ another ${\gamma}(1)$ bits changed to $|1\ran$, 
etc., up to step ${k}+1+\tau$, when ${\gamma}(\tau)$ bits flipped to 
$|1\ran$.

\paragraph*{Remark 1:} 

Note, that elements of ${\Gamma}(N,\tau)$ do not depend on $k$. The 
same $\gamma$ describes many different rounds of 
${\cal{G}}_{(N,\tau)}$, as they might
 correspond to  different values of $k$.
${\scriptscriptstyle{\square}}$

\subsection{Notation}\label{section3.1}

Every set ${\Gamma}(N,\tau)$ can be divided into subsets collecting
rounds during which the same number of bits changed from $|0\ran $
to~$|1\ran$. Therefore, we introduce:
\ba
\forall {\mathrm{M}}\in (\overline{1,N})\;\;\;\;\Gamma_{\mathrm{M}}(N,\tau)\defeq \left\{{\gamma}\in {\Gamma}(N,\tau)
: \sum_{j=0}^\tau {\gamma}(j)={\mathrm{M}}
\right\}.
\n 
\ea
Let us fix the parameters $(N,\tau)$ and M. Set $\Gamma_{\mathrm{M}}(N,\tau)$ 
consists of all these elements of ${\Gamma}(N,\tau)$ in which
between steps ${k}+1$ and ${k}+1+\tau$ exactly ${\mathrm{M}}$ bits
flipped  to state~$|1\ran$. Obviously,
\ba
{\Gamma}(N,\tau)=\bigcup_{{\mathrm{M}}=1}^N \Gamma_{\mathrm{M}}(N,\tau)= \Gamma_N (N,\tau)\cup
\left( \bigcup_{  {\mathrm{M}}=1}^{N-1} \Gamma_{\mathrm{M}}(N,\tau) \right).\n
\ea
To explicitly distinguish between
successful and unsuccessful events, we define:
\ba
{\mathrm{S}}(N,\tau)\defeq\Gamma_N (N,\tau)\quad {\mathrm{and}}\quad
{\mathrm{F}}(N,\tau)\defeq\bigcup_{{\mathrm{M}}=1}^{N-1}
\Gamma_{\mathrm{M}}(N,\tau). \label{eq3_7}
\ea
The set of successful events, ${\mathrm{S}}(N,\tau)$, consists of
these events in which, between steps ${k}+1$ and ${k}+1+\tau$, all
$N$ bits flipped to $|1\ran$.  The complementary set,
${\mathrm{F}}(N,\tau)$, consists of all events in which after step
${k}+1+\tau$ at least one bit was still left in state $|0\ran$.
These events will be referred to as unsuccessful, or resulting in a
failure. For further convenience, we  single out subsets of these
elements of ${\Gamma}(N,\tau)$ for which the last from the bit-flips
occurred exactly at step ${k}+1+\s$:
\ba
\fl
\forall \sigma\in (\overline{0,\tau})\qquad\Gamma^\s(N,\tau)\defeq \left\{{\gamma}\in {\Gamma}(N,\tau):
 {\mathrm{max}} \{ l \in(\overline{0,\tau}): {\gamma}(l)>0 \}=\s
\right\} .\n
\ea
Similarly, we define sets $\Gamma_{\mathrm{M}}^\s(N,\tau)$ as the
intersection of $\Gamma_{\mathrm{M}}(N,\tau)$ and $
\Gamma^\s(N,\tau)$:
\ba
\forall ({\mathrm{M}},\s)\in (\overline{1,N})\times(\overline{0,\tau})
\qquad\Gamma_{\mathrm{M}}^\s(N,\tau)\defeq
\Gamma_{\mathrm{M}}(N,\tau)\cap \Gamma^\s(N,\tau).\n
\ea
In particular,  a set of these successful events for which the
last bit-flip occurred at step $k+1+\s$ is denoted by $
\Gamma_{N}^\s(N,\tau)$. Note, that the sum of $
\Gamma_{\mathrm{M}}^\s(N,\tau)$ over all
${\mathrm{M}}\in(\overline{1,N})$ and $\s\in(\overline{0,\tau})$
equals to $\Gamma(N,\tau)$.

\subsection{Measures $g_{\scriptscriptstyle{
(N,}\tau {\scriptscriptstyle{)}}}$, numbers $sp(n,\nu)$}

To be able to write down calculations in more condensed form, we
introduce  the  maps: {\mbox{$\;
\,\mathbf{\widetilde{}}\,,\;\mathbf{\bar{}}\, \in \mathrm{Map}\left({\Gamma}(N,\tau),
{\mathrm{Map}}\left((\overline{0,\tau}),(\overline{0,N})\right)\right)
$}}, defined by
\ba
\ww(l)\!\defeq\!  N-\sum_{j=0}^{l-1}\w(j)\quad
\mathrm{and}\quad
 \bar\w(l)\!\defeq\!\ww(l)-\w(l),
 \n
\ea
for $l \in (\overline{0,\tau})$. For a given 
$\gamma\in{\Gamma}(N,\tau)$,   values $\mathbf{\tilde{\w}}(l)$ and 
$\mathbf{\bar{\w}}(l)$ are equal to the number of bits in states 
$|0\ran$ at steps ${k}+l$ and  ${k}+l+1$, respectively. It is easy 
to check that $\bar\w(\tau)=N-\sum_{j=0}^\tau
\w(j)$ and that, for
$l\in(\overline{0,\tau\!-\!1})$, the identity $\bar\w (l)=\ww(l+1)$
holds. Using this notation, we introduce maps
${g}_{\scriptscriptstyle{ (N,}\tau {\scriptscriptstyle{)}}}$ as 
follows:
\ba
{g}_{\scriptscriptstyle{ (N,}\tau {\scriptscriptstyle{)}}}:
{\Gamma}(N,\tau)
 \ni {\gamma} \longrightarrow  {g}_{\scriptscriptstyle{ (N,}\tau {\scriptscriptstyle{)}}}(\w)
\defeq \prod_{l=0}^\tau {\ww(l) \choose \w(l)} p^{\w(l)}q^{\bar\w(l)}
\in[0,1]\,.\label{measure_g}
\ea
Measures on ${\Gamma}(N,\tau)$, defined as the extensions of
\eref{measure_g}, are not normalized to 1; we
have 
\ba
\sum_{\scriptstyle{\w \in{\Gamma}(N,\tau)} }
{g}_{\scriptscriptstyle{ (N,}\tau
{\scriptscriptstyle{)}}}(\w)=1-q^N\,,\quad
\n
\ea
which reflects the fact that events `at step
${k}+1$ all bits remain in state $|0\ran$' do not belong to
${\Gamma}(N,\tau)$.

\paragraph*{Remark 2:}
Values of ${g}_{\scriptscriptstyle{ (N,}\tau
{\scriptscriptstyle{)}}}(\w)$ cannot be interpreted as
probabilities characterizing~$\w$; pairs
$({\Gamma}(N,\tau),{g}_{\scriptscriptstyle{ (N,}\tau
{\scriptscriptstyle{)}}})$ do not define probability spaces.
$\quad {}_{\scriptscriptstyle{\square}}$
 \\

\noindent
For clarity of notation, we introduce functions $sp(n,\nu)$
defined, for $n\geq 1$ and $\nu\geq 0$, by
\ba
sp(n,\nu)\defeq (1-q^{\nu+1})^n-q^n(1-q^\nu)^n \label{sp} \;.
\ea
Additionally, we assume that $\,sp(n,-1)\defeq 0\,$. Various
combinations of numbers $sp(n,\nu)$ appear repeatadly in the
calculations of measures $g_{\scriptscriptstyle{ (N,}\tau
{\scriptscriptstyle{)}}}$ for the important subsets of
$\Gamma(N,\tau)$. In
\ref{A_B}, it is shown that:
\ba
{g}_{{\scriptscriptstyle{ (N,}}\tau 
{\scriptscriptstyle{)}}}({\mathrm{S}}(N,\tau))=
sp(N,\tau), \label{SN}\\
g_{ {\scriptscriptstyle{ (N,}}\tau {\scriptscriptstyle{)}}}
({\mathrm{F}}(N,\tau) )
=1-q^N-sp(N,\tau), \label{FN} \\
{g}_{{\scriptscriptstyle{ (N,}}\tau
{\scriptscriptstyle{)}}}(\Gamma_N^{\s}(N,\tau))= sp(N,\s)-sp(N,\s-1),
\label{Ss}
\ea
for any $\s\in\left(\overline{0,\tau}\right)$. Comparing  (\ref{SN})
and (\ref{Ss}) we see that, as expected, values of measures
corresponding to successes that occurred exactly at a step $k+1+\s$
is equal to the values of measures corresponding to the sum of all
successes up to the step $k+1+\s$ minus the values of measures
corresponding to sum of all successes up to the step $k+\s$.

\subsection{Expected sum values $\lan\lan \,\lambda_{{\scriptscriptstyle{ (N,}}\tau
{\scriptscriptstyle{)}}}\,\ran\ran $}\label{section3_3}

Let  functions $\,\lambda_{{\scriptscriptstyle{ (N,}}\tau
{\scriptscriptstyle{)}}}$ be equal to $\tau$ for all unsuccessful
events, and equal to $\s$ for events that ended successfully at step 
$k+1+\s$, i.e., let $\,\lambda_{{\scriptscriptstyle{ (N,}}\tau 
{\scriptscriptstyle{)}}}\in
\rm{Map}({\Gamma}(N,\tau),{\mathbb{R}})$ be defined by $\lambda_{{\scriptscriptstyle{ (N,}}\tau
{\scriptscriptstyle{)}}}({\mathrm{F}}(N,\tau) ) \defeq \{\tau \} $ 
and $\lambda_{{\scriptscriptstyle{ (N,}}\tau 
{\scriptscriptstyle{)}}}(\Gamma_N^\s(N,\tau)) \defeq \{\s \}$, for 
$\s\in\left(\overline{0,\tau}\right)$. We want to calculate expected 
sum values~$\lan\lan
\,\lambda_{{\scriptscriptstyle{ (N,}}\tau
{\scriptscriptstyle{)}}}\,\ran\ran$  over the sets of unsuccessful
events ${\mathrm{F}}(N,\tau)$ and the sets of successful events
${\mathrm{S}}(N,\tau)$. To shorten  the formulas, from now on, we
will write $g,\,\lambda$ instead of  ${g}_{{\scriptscriptstyle{
(N,}}\tau {\scriptscriptstyle{)}}},\,\lambda_{{\scriptscriptstyle{
(N,}}\tau {\scriptscriptstyle{)}}} $,  whenever omission of indexes
does not lead to confusion. From~\eref{FN}, it is clear that:
\ba
\lan\lan \,\lambda\,\ran\ran_{ \left( \mathrm{F}(N,\tau),g \right) }\defeq
\tau {g} \left( {\mathrm{F}}(N,\tau) \right)
=\tau\left(1-q^N-sp(N,\tau)\right).\label{l22}
\ea
Similar relation for sets $S(N,\tau)$, given by
\ba
\lan\lan \,\lambda\,\ran\ran_{ \left( \mathrm{S}(N,\tau),g \right) } & \defeq &
\sum_{\sigma=0}^{\tau}\sigma {g} \big(
\Gamma_N^{\s}(N,\tau)\big)= \tau sp(N,\tau)- \sum_{j=0}^{\tau-1} sp(N,j)\,, \label{e14}
\ea
is derived  in \ref{A_B4}. Combining \eref{l22} and \eref{e14} we 
obtain the expected sum values corresponding to $\Gamma(N,\tau)$:
\ba
\fl
\lan\lan \,\lambda \, \ran\ran_{ \left( \Gamma(N,\tau) , {g} \right) }
& \defeq& \sum_{\w\in{\Gamma}(N,\tau)}
\lambda(\w){g}(\w)=\sum_{\w\in{\mathrm{F}}(N,\tau)}
\lambda(\w){g}(\w)
+\sum_{\w\in{\mathrm{S}}(N,\tau)}
\lambda(\w){g}(\w)=
\n\\
&=&\tau {g}\left( {\mathrm{F}}(N,\tau) \right) +
\sum_{\s=0}^{\tau} \s {g}\left(\Gamma_N^\s(N,\tau)\right)
=\tau(1-q^N)-\sum_{j=0}^{\tau-1} sp(N,j). \n
\ea

\paragraph{Remark 3:}\label{remark3}
Because pairs $( {\Gamma(N,\tau)} , {g} ) $ are not  probability
spaces, functions  $\lambda$ are not  random variables, and
$\lan\lan\,
\lambda \,\ran\ran_{ \left( \Gamma(N,\tau) , {g} \right) }$ cannot
be interpreted as the average values.
 (See Remark~2 and  \ref{A_A1}.) $\;{\scriptscriptstyle{\square}}$

\section{Single rounds of ${\cal{G}}_{(N,\tau)}$ }\label{section4}
\subsection{Sample spaces $ {\mathbb{B}}(N,\tau)$}
A description characterizing  single rounds  of
${\cal{G}}_{(N,\tau)}$  from the beginning -- not only the second
parts as it was done in previous section --  has to include
information about all steps preceding the first bit-flips.
In order to do so, we define  sets $ {\mathbb{B}}(N,\tau)\defeq
{\mathbb{N}}\times {\Gamma}(N,\tau).$ Let us fix parameters
$k,\,\tau,\,N\in\mathbb{ N}$, for $N\geq 2$, and choose some
$\gamma\in {\Gamma}(N,\tau)$.  The element $(k,\w)\in
{\mathbb{B}}(N,\tau)$ corresponds to the round in which $N$ bits
remained in states~$|0\ran$ for initial $k$ steps, then  $\w(0)$
bits was transferred to~$|1\ran$ at step $k+1$, $\w(1)$ bits flipped
to~$|1\ran$ at step $k+2$, etc.. Additionally, we define maps
$\B_{{\scriptscriptstyle{ (N,}}\tau {\scriptscriptstyle{)}}}$
\ba
 \beta_{{\scriptscriptstyle{ (N,}}\tau
{\scriptscriptstyle{)}}}:{\mathbb{B}}(N,\tau)\ni
(k,\w)\longrightarrow \beta_{{\scriptscriptstyle{ (N,}}\tau
{\scriptscriptstyle{)}}}(k,\w)\defeq q^{Nk}  g_{{\scriptscriptstyle{
(N,}}\tau {\scriptscriptstyle{)}}}(\w)
\in[0,1]\,.
 \label{mapaB}
\ea
Extensions of \eref{mapaB}  introduce measures on the corresponding
sets ${\mathbb{B}}(N,\tau)$. It is easy to check the normalization:
\ba
\B_{{\scriptscriptstyle{ (N,}}\tau
{\scriptscriptstyle{)}}}( {\mathbb{B}}(N,\tau) ) &=&\!\!\!\!\!\!
\sum_{ (k,\w)\in{\mathbb{B}}(N,\tau) } \!\!\!\!\!\!
\B_{{\scriptscriptstyle{ (N,}}\tau
{\scriptscriptstyle{)}}}(k,\w) =\sum_{k=0}^{\infty} q^{Nk}
\!\!\!
\sum_{\w\in \Gamma(N,\tau)}
\!\!\!\!
g_{{\scriptscriptstyle{ (N,}}\tau {\scriptscriptstyle{)}}}(\w)=1,
\label{normB}
\ea
and conclude that pairs $(
{\mathbb{B}}(N,\tau),\B_{{\scriptscriptstyle{ (N,}}\tau
{\scriptscriptstyle{)}}})$ define probability spaces.

Every  set $ {\mathbb{B}}(N,\tau) $ splits into a subset of rounds
that ended successfully, i.e., all $N$ bits were transferred to
states~$|1\ran$, and a complementary set of rounds that ended in
failure. We denote these subsets ${\mathbb{S_B}}(N,\tau)$ and
${\mathbb{F_B}}(N,\tau)$, respectively, and using \eref{eq3_7} write
them as
\ba
{\mathbb{S_B}}(N,\tau)\defeq
{\mathbb{N}}\times{\mathrm{S}}(N,\tau) \quad {\mathrm{and}} \quad
{\mathbb{F_B}}(N,\tau)\defeq     {\mathbb{N}}\times
{\mathrm{F}}(N,\tau) .\n
\ea
Taking relations (\ref{sp}), (\ref{SN}) and (\ref{FN}) into
account we obtain after a brief calculation:
\ba
\B_{{\scriptscriptstyle{ (N,}}\tau
{\scriptscriptstyle{)}}}\big({\mathbb{S_B}}(N,\tau)
\big)=sp(N,\tau)/(1-q^N)\eqdef {\cal{P}}(N,\tau),\label{defP}
\\
 \B_{{\scriptscriptstyle{ (N,}}\tau
{\scriptscriptstyle{)}}}\big({\mathbb{F_B}}(N,\tau)
\big) =1-sp(N,\tau)/(1-q^N)\eqdef {\cal{Q}}(N,\tau).\label{defQ}
\ea
Definitions of ${\cal{P}}(N,\tau)$ and ${\cal{Q}}(N,\tau)$,
introduced above, will be used to shorten formulas in the sections
that follow.

\subsection{Average number of steps in a single round of ${\cal{G}}_{(N,\tau)}$,
 i.e., $\lan\, \Lambda\, \ran_{({\mathbb{B}}(N,\tau),\B)}$}

The number of steps in any given round $(k,\w)$ is equal to the sum
of $k$ steps before the first bit-flip plus 1 (the step at which
this first transfer from $|0\ran$ to $|1\ran$ occurred)  plus
$\lambda(\w)$ steps after the first bit-flip. The corresponding
random variables on sets ${\mathbb{B}}(N,\tau)$ are defined as
\ba
\Lambda_{{\scriptscriptstyle{ (N,}}\tau {\scriptscriptstyle{)}}}:
{\mathbb{B}}(N,\tau)\ni(k,\w)\longrightarrow
\Lambda_{{\scriptscriptstyle{ (N,}}\tau {\scriptscriptstyle{)}}}({k},\w)
={k}+1+\lambda_{{\scriptscriptstyle{ (N,}}\tau
{\scriptscriptstyle{)}}}(\w)\in {\mathbb{R}}.\n
\ea
 Using (\ref{l22}), (\ref{e14}), and dropping 
indexes $\scriptstyle{(N,\tau)}$, we calculate the expected sum 
values $\lan\lan\,
\Lambda\, \ran\ran$
over sets $ {\mathbb{S_B}}(N,\tau)$ and  ${\mathbb{F_B}}(N,\tau)$ 
and obtain: 
\ba
\lan\lan\, \Lambda\, \ran\ran_{({\mathbb{S_B}}(N,\tau),\B)}\defeq\sum_{(k,\w)\in
{\mathbb{S_B}}(N,\tau)}\!\!\!\!\!\!
\Lambda(k,\w)\B(k,\w)=\n\\
= g_{}(\mathrm{S}(N,\tau))(1-q^N)^{-2}  +
\lan\lan\, \lambda \,\ran\ran_{ \left({\mathrm{S}}(N,\tau) , {g} \right) }(1-q^N)^{-1} \n\\
=\Big(\big[(1-q^N)^{-1} +\tau\big]\cdot {sp(N,\tau)}
-\sum_{j=0}^{\tau-1} sp(N,j)\Big) (1-q^N)^{-1}
\label{L19}
\ea
and
\ba
\lan\lan \,\Lambda \,\ran\ran_{({\mathbb{F_B}}(N,\tau),\B)}\defeq
\sum_{(k,\w)\in
{\mathbb{F_B}}(N,\tau)}\!\!\!\!\!\!
  \Lambda(k,\w) \B(k,\w) =\n\\
= g(\mathrm{F}(N,\tau))(1-q^N)^{-2}+ \tau
g(\mathrm{F}(N,\tau))(1-q^N)^{-1}
\n\\
=  g(\mathrm{F}(N,\tau)) \left[
\tau+(1-q^N)^{-1} \right](1-q^N)^{-1}\n\\
=  \left[
\tau+(1-q^N)^{-1} \right] {\cal{Q}}(N,\tau)\label{L20}\,,
\ea
respectively. The average number of steps in single rounds of 
${\cal{G}}_{(N,\tau)}$ can be calculated as the sum of (\ref{L19}) 
and~(\ref{L20}), which yields:
\ba
\lan\, \Lambda\, \ran_{({\mathbb{B}}(N,\tau),\B)}=
\lan\lan\, \Lambda\, \ran\ran_{({\mathbb{S_B}}(N,\tau),\B)}
+\lan\lan\, \Lambda\, \ran\ran_{({\mathbb{F_B}}(N,\tau),\B)}\n\\
=\tau+\Big(1   -\sum_{j=0}^{\tau-1} sp(N,j)\Big)\big(1-q^N \big)^{-1}\n\\
= \tau+\Big(1-\big(1-q^\tau\big)^N \Big)\big(1-q^N \big)^{-1} -
\sum_{j=1}^{\tau-1} (1-q^j)^N
\,.
\label{L21}
\ea
This formula will be essential for  calculations of the average
number of steps in the whole game. Due to (\ref{normB}), \eref{L21}
corresponds to the real average value, i.e.,  the mean or expected 
value in the standard sense; for  differences between the notations 
$\lan\lan
\,\cdot\,\ran\ran$ and $\lan
\,\cdot \,\ran$  see
\ref{A_A1}.

\subsection{Order in sets ${\mathrm{S}}(N,\tau)$ and  $ {\mathrm{F}}(N,\tau) $}

It is convenient to label elements of a given $\Gamma(N,\tau)$,
indexing separately elements in the subsets of successfully and
unsuccessfully ended rounds, i.e., the elements in
${\mathrm{S}}(N,\tau)$ and ${\mathrm{F}}(N,\tau)$. Because $
\,|\Gamma(N,\tau)|\leq N^{\tau+1}\,
$, all subsets of $\Gamma(N,\tau)$ are finite. Consequently, there
exist natural numbers $I$ and $J$ such that
\ba
|{\mathrm{S}}(N,\tau)| \eqdef I(N,\tau)=I \quad{\mathrm{ and}} \quad
| {\mathrm{F}}(N,\tau) | \eqdef J(N,\tau)=J. \n 
\ea
We choose two arbitrary (but fixed for further considerations)
bijective functions:
 $ \varphi_{{\scriptscriptstyle{ (N,}}\tau
{\scriptscriptstyle{)}}} \in \rm{Map}(
(\overline{1,I}),{\mathrm{S}}(N,\tau)) $ and $
\psi_{{\scriptscriptstyle{ (N,}}\tau {\scriptscriptstyle{)}}}
 \in \rm{Map}((\overline{1,J}),{\mathrm{F}}(N,\tau)) $.
Then, for a given $(N,
\tau)$ and every
$i
\in (\overline{1,I})$,
$j
\in (\overline{1,J})$,
we denote
 $\varphi_{{\scriptscriptstyle{ (N,}}\tau
{\scriptscriptstyle{)}}}(i)\eqdef s_i$  and
$\psi_{{\scriptscriptstyle{ (N,}}\tau
{\scriptscriptstyle{)}}}(j)\eqdef f_j\,$, which allows to express the
sets ${\mathrm{S}}(N,\tau)$ and ${\mathrm{F}}(N,\tau)$ as lists of
the ordered elements:
\ba{\mathrm{S}}(N,\tau)=\{s_1, s_2,\dots,s_I\}
\quad \mathrm{and} \quad
 {\mathrm{F}}(N,\tau)=\{f_1,f_2,\dots,f_J\}.\n
 \ea
This notation is easily extended to the following subsets of
${\mathbb{B}}(N,\tau)$
\ba
{\cal{S}}_i={\cal{S}}_i(N,\tau) \defeq {\mathbb{N}}\times\{s_i\}
\quad{\mathrm{and}}\quad {\cal{F}}_j={\cal{F}}_j(N,\tau) \defeq
{\mathbb{N}}\times\{f_j\} \,,\n
\ea
where, again, $i \in \left(\overline{1,I}\right)$ and $j \in
\left(\overline{1,J}\right)$. 
The probabilities  corresponding to ${\cal{S}}_i$ and ${\cal{F}}_j$ 
are equal to:  
\ba
\B({\cal{S}}_i)= g(s_i) (1-q^N)^{-1}\n
\quad {\mathrm{
and}}
\quad
\B({\cal{F}}_j)= g(f_j)(1-q^N)^{-1}\,. \label{e35}
\ea

\section{Games ${\cal{G}}_{(N,\tau)}$}\label{section5}

In this section,  a general description of the whole course of an
arbitrary game ${\cal{G}}_{(N,\tau)}$  (starting with all $N$ bits
of the system in state $|0\ran$  and ending with all bits in state
$|1\ran$) is presented. To shorten the formulas,  elements of a
multi-indexes notation are used; for details of the notation see
{\mbox{\ref{A_A1}}}.

\subsection{Sample spaces $ {\mathbb{D}}(N,\tau)$}

Let $ N\geq 2,\,\tau\geq 0$ be fixed natural numbers, and $J
\defeq | {\mathrm{F}}(N,\tau) |$. We define  a family of sets $Y_{\mathrm{M}}(N,\tau)$
specified, for $\mathrm{M} \in \mathbb{ N}$, by the formula:
\ba
\fl
Y_{\mathrm{M}}(N,\tau) &\defeq& \Big[{\mathrm{Map}}\big(
\big({\overline{0,{\mathrm{M}}}}\big),{\mathbb{N}}\big)\times
\Big\{\zeta\in {\mathrm{Map}}\big( \big({\overline{1,J}}\big),{\mathbb{N}}\big):
|\zeta|={\mathrm{M}} \Big\}\Big] \times S(N,\tau) .\n
\ea
Next, with every family of sets $Y_{\mathrm{M}}(N,\tau)$ we 
associate the corresponding sample space $ {\mathbb{D}}(N,\tau)\defeq
\bigcup_{{\mathrm{M}}=0}^{\infty} Y_{\mathrm{M}}(N,\tau).$ 
Elements of $ {\mathbb{D}}(N,\tau)$ correspond to all possible
courses of game~${\cal{G}}_{(N,\tau)}$. To be more specific:
$(\xi,\zeta,\gamma)\in {\mathbb{D}}(N,\tau)$ denotes a course of 
game ${\cal{G}}_{(N,\tau)}$, which started with $|\zeta|$ 
unsuccessful rounds  and ended with successful round characterized 
by $(k,\gamma)$, where  $k=\xi(0)$. Values of $\zeta(j)$, for $j\in 
({\overline{1,J}})$, match number of times that unsuccessful rounds 
of  type  $f_j\,$ were repeated. Number of steps during first parts 
of rounds that ended unsuccessfully are given by $\xi(l)$, for 
$l\in({\overline{1,|\zeta|}})$.

We introduce also measures on sets ${\mathbb{D}}(N,\tau)$,  defining 
them as extensions of maps ~$\delta_{{\scriptscriptstyle{ (N,}}\tau 
{\scriptscriptstyle{)}}} \in \rm{Map}({\mathbb{D}}(N,\tau),[0,1])$ 
given by
\ba
\delta(\xi,\zeta,\gamma)\defeq
{g}(\gamma) \frac{|\z|!}{\z!} \prod_{l=0}^{|\z|} q^{N
\xi(l)}\prod_{j=1}^J \left(g(f_j)^{\zeta(j)} \right). \n 
\ea
As before, we dropped indexes $\scriptstyle{(N,\tau)}$ referring to 
measures $g$, $\delta$ or $\B$. It is relatively easy to show that:
\ba
\delta
\big( {\mathbb{D}}(N,\tau) \big)= \sum_{(\xi,\z,\gamma)\in{\mathbb{D}}(N,\tau) }
\delta(\xi,\zeta,\gamma)=1\,,\label{nD}
\ea
which means that pairs $({\mathbb{D}}(N,\tau),\delta)$ define
probability spaces. Derivation of (\ref{nD}) is presented in
\ref{A_C2}.

\subsection{Average number of steps  $\lan\, {\mathcal{K}}\,\ran_{({\mathbb{D}}(N,\tau),\delta)}$}

We introduce random variables $ {\cal{K}}_{{\scriptscriptstyle{
(N,}}\tau {\scriptscriptstyle{)}}} \in 
\rm{Map}({\mathbb{D}}(N,\tau),{\mathbb{R}})  $,  defined by
\ba
  {\cal{K}}_{{\scriptscriptstyle{ (N,}}\tau
{\scriptscriptstyle{)}}}(\xi,\zeta,\gamma)\defeq
      |\xi|+(\tau+1)|\zeta|+\lambda(\gamma)+1
. \label{Kdef}
\ea
For fixed $(N,\tau)$, the value  $ {\cal{K}}_{{\scriptscriptstyle{
(N,}}\tau {\scriptscriptstyle{)}}}(\xi,\zeta,\gamma)$   equals to 
the number of steps in the course of game ${\cal{G}}_{(N,\tau)}$ 
characterized by $(\xi,\zeta,\gamma)$. Terms building \eref{Kdef} 
can be easily interpreted: all $N$ bits are in initial state 
$|0\ran$ for $|\xi|$ steps; a length of the second part of 
successful round equals to $\lambda(\gamma)+1$; a length of the 
second part of every unsuccessful round is given by $(\tau+1)$, and 
$|\zeta|$ equals to the number of unsuccessful rounds.

In \ref{A_C3}, it is shown that  average number of steps $\lan\,
{\mathcal{K}}\,
\ran_{{\mathbb{D}}(N,\tau)}$ is equal to:
\ba
\lan\, {\mathcal{K}}\, \ran_{({\mathbb{D}}(N,\tau),\delta)} \defeq
\!\!\!\!\!\!\!
\sum_{(\xi,\z,\gamma)\in{\mathbb{D}}(N,\tau) }
\!\!\!\!\!\!\!
{\cal{K}}(\xi,\zeta,\gamma)\delta(\xi,\zeta,\gamma)=
 \lan\, \Lambda\, \ran_{({\mathbb{B}}(N,\tau),\B)}
{\cal{P} }(N,\tau)^{-1}  \;.\label{Kavr}
\ea
Using  (\ref{defP}) and  (\ref{L21})  we finally obtain:
\ba
\lan\, {\mathcal{K}}\, \ran_{({\mathbb{D}}(N,\tau),\delta)}=
 \label{nowe3}\\
\fl \n
\Big[(1-q^\tau)^N -1 -\tau(1-q^N)+ (1-q^N) \sum_{j=1}^{\tau-1}   (1-q^j)^N
\Big] \Big[q^N(1-q^\tau)^N-(1-q^{\tau+1})^N  \Big]^{-1}\,,
\ea
i.e., exactly the formula proclaimed earlier in
section~\ref{section2.2}. Note, that although the right-hand side of
(\ref{nowe3}) is well defined
 for $N\in
\mathbb{R}$, the left-hand side  was derived with the
assumption that $N \geq 2$ is a natural number. Similarly, $\lan\,
{\mathcal{K}}\, \ran_{({\mathbb{D}}(N,\tau),\delta)}$  rewritten
as in (\ref{inne3})  has a form of the function  well defined
 for $\tau\in \mathbb{R}$, however, it was derived only for natural  $\tau$. 

For $ \tau=0$, the average number of steps is equal to
{\mbox{$\lan\, {\mathcal{K}}\,
\ran_{({\mathbb{D}}(N,0),\delta)}=(1-q)^{-N}=p^{-N}\,$}}. The limit of $
\lan\, {\mathcal{K}}\, \ran_{({\mathbb{D}}(N,\tau),\delta)}$ for
$\tau\rightarrow\infty$ is calculated in \ref{AppLimit}; it is given
by:
\ba
\lim_{\tau\rightarrow\infty} \lan\, \mathcal{K}\, \ran_{({\mathbb{D}}(N,\tau),\delta)}=
\sum_{k=1}^{N}(-1)^{k+1} {N \choose k} \frac{1}{1-q^N} \;.\label{limit}
\ea
As expected, (\ref{limit}) is identical to (\ref{perfect}) because 
this limit corresponds to the perfect memory case. We should also 
note that \eref{nowe3} can be used to approximate values of
\eref{limit}, as for large $N$ its form is far more convenient for 
numerical calculations. Analysis of accuracy of this approximation 
and comparison with results from 
\cite{Eisenberg} is an interesting problem, which we plan to address in the near future.

\subsection{Intuitive interpretation of (\ref{Kavr})
and a non-zero reset time}\label{section5_3} A formal derivation of 
(\ref{Kavr}) involved some calculations, but the result itself is 
not surprising. In the simplest case ($N=1$) of game~${\cal{G}}_N$, 
described in section~\ref{section2.1},  the average number of steps 
needed to obtain the success is equal to
\ba
 \lan \mathcal{K}_{{\cal{G}}_1} \ran =\frac{1}{p} \;.\n
\ea
If procedure $T$ with probability $p$ transfers a bit from state
$|0\ran$  to  $|1\ran$, then to actually flip this bit from $|0\ran$
to $|1\ran $ on average $1/p$ applications of procedure $T$ are
need. If time needed  for a single application of $T$ is equal to
$\lan t_1 \ran$, then the
 average waiting time for the  successful transfer is given by
\ba
\Big( {\mathit{time\;needed\;for}}\; \lan \mathcal{K}_{{\cal{G}}_1} \ran\Big)
=\frac{\lan t_1
\ran}{p}
\;.\label{gN1}
\ea
Note, that (\ref{Kavr})  has exactly the same structure as
(\ref{gN1}). Average number of steps needed for the successful end 
of the total game~${\cal{G}}_{(N,\tau)}$ equals to the duration 
(average number of steps) of one round divided by the probability of 
success in a single round. Another simple model connected to this 
intuition is presented in \ref{A_E}.

Up to now, we were assuming that after an unsuccessfully ended round
the system is immediately reset to its initial state, but it is easy
to release this assumption. To include a non-zero reset time
$\Omega$, where $\Omega$ is also measured in steps defined by a time
needed for application of procedure $T$ to the system considered, we
just define the new random variables: $   
{\cal{K}}_\Omega(\xi,\zeta,\gamma)\defeq
       {\cal{K}}(\xi,\zeta,\gamma)+\Omega |\zeta|
.$ Using  \eref{component2}, we obtain the formula: 
\ba
  \lan\, \mathcal{K}_\Omega\,
\ran_{({\mathbb{D}}(N,\tau),\delta)}=
      \lan\, \mathcal{K}\,
\ran_{({\mathbb{D}}(N,\tau),\delta)}+ \Omega \mathcal{Q}(N,\tau) \mathcal{P}(N,\tau)^{-1}\
. \n
\ea
Again, this result has very simple interpretation: on average, there 
are $\mathcal{Q}(N,\tau)
\mathcal{P}(N,\tau)^{-1}$ unsuccessfully ended rounds for
one round that ended successfully.
 Thus, the term $\Omega\,
\mathcal{Q}(N,\tau) \mathcal{P}(N,\tau)^{-1}$ that has to be added to
$\lan\, \mathcal{K}\,
\ran_{({\mathbb{D}}(N,\tau),\delta)}$ to obtain $\lan\,\mathcal{ K}_\Omega\,
\ran_{({\mathbb{D}}(N,\tau),\delta)}$.

\section*{Summary}
We have studied the system of $N$ independent identical bits 
(memories) and derived the formula for the average waiting time for 
obtaining $N$ successful bit-flips under the condition that the time 
difference between the first and the last from the bit-flips is not 
longer then $\tau$. Using simple mathematical tools we have 
provided  an accurate, analytical description of this system and 
presented the results in a form allowing for effective numerical 
calculations for arbitrary parameters $N$ and  $\tau$. Connecting 
the result obtained with a quantum repeater rate analysis in the 
case of imperfect memories, we have shown that it can be also used 
as an approximation  of identity characterizing the perfect memory 
case -- a problem that was studied not only in the context of 
quantum repeater~\cite{Nadja1}, but also in bioinformatics  
\cite{Eisenberg}.

\ack
This work was partially supported by the German Ministry of 
Education and Research grant QuOReP and the European Commission 
STREP-project MALICIA.

\appendix
\section{Notation and Conventions. }\label{A_A}
\setcounter{section}{1}
\subsection{Basic notation}\label{A_A1}
$ {\mathbb{N}}=\{0,1,2,\dots\,\} $ denotes set of natural numbers
including 0. For $a,b\in{\mathbb{N}}$ and $a \leq b$ we define
$\left({\overline{a,b}} \right)\defeq \{n\in{\mathbb{N}}: a\leq n
\leq b \} $.
Set of real numbers is denoted by  ${\mathbb{R}}$.
For
$r,s\in{\mathbb{R}}$ and $r\leq s$ we define
$[r,s]\defeq \{z\in{\mathbb{R}}: r\leq z \leq s \} $ and
$\, ]r,s[\,\defeq \{z\in{\mathbb{R}}: r < z < s \} $.\\
We assume that for $a\in {\mathbb{N}}$,  $\,f:
{\mathbb{N}}\longrightarrow{\mathbb{R}}\,$, and an integer number
$b$ from inequality $a>b\,$ follows that
\ba
\sum_{n=a}^b f(n) \defeq 0 \qquad
{\mathrm{and}} \qquad
\prod_{n=a}^b f(n) \defeq 1.\n
\ea
For sets B and ${\mathrm{A}}\neq
\emptyset\;$, the  set of all mappings of A into B is denoted by Map(A,B).
For any map {\mbox{$\zeta\in
{\mathrm{Map}}\left(\left(\overline{a,b}\right),{\mathbb{N}}\right)$}}
we define $|\zeta|$ and $ \zeta!\, $ as
\ba
|\zeta| \defeq \sum_{n=a}^b \zeta(n)
\qquad {\mathrm{and}}
\qquad
\zeta!\defeq\prod_{n=a}^b \left(\zeta(n)\right)!\,. \n
\ea
In this notation
\ba
\frac{|\zeta|!}{\zeta!}=\frac{\big(\zeta(a)+\zeta(a+1)+\dots+\zeta(b)    \big)!}{\big(
\zeta(a)!\big)\big( \zeta(a+1)!\big)\n
 \dots \big(\zeta(b)!\big)}\,. \\ \n
\ea
Let $\mathbb{M}$ be a finite (or countable) set and $\mu$ a map
 $\mu:{\mathbb{M}}\longrightarrow [0, 1]\,$;  
$({\mathbb{M}},\mu)$  denotes the space $\mathbb{M}$ with measure
$\mu$ defined on a $\s$-algebra $2^{\mathrm{M}}$ of all subsets of
$\mathbb{M}$:
\ba
\mu:2^{\mathbb{M}}\ni \mathrm{A}\longrightarrow \mu(A)\defeq
\cases{0 \qquad\qquad\; \mathrm{A}=\emptyset\\
      \sum_{\mathrm{a}\in \mathrm{A}} \mu(\mathrm{a})\;\;\quad\mathrm{ A}\neq \emptyset
       }  \quad \in (\mathbb{R}\cup \{\infty\}).\n
\ea
All pairs $({\mathbb{M}},\mu)$ considered in this  paper fulfil
condition $\mu(\mathbb{M})\leq 1$. For clarity, cases
$\mu({\mathrm{A}}) < 1$
 and $  \mu({\mathrm{A}}) = 1$, for $\mathrm{A} \subseteq \mathbb{M}$, are treated separately.
 To distinguish between them,  for a given map
 $ f:{\mathbb{M}}  \longrightarrow  \mathbb{R}$, a value of the sum
$\sum_{\mathrm{a} \in {\mathrm{A}}} f(\mathrm{a})\mu(\mathrm{a})$ is
denoted in this paper by
\ba
\lan\lan\, f\,
\ran\ran_{({\mathrm{A}},\mu)} \defeq & \sum_{\mathrm{a} \in {\mathrm{A}}} f(\mathrm{a})\mu(\mathrm{a})
\qquad \mathrm{when } \qquad\mu({\mathrm{A}}) < 1 \n
\\
\fl{\mathrm{and\; by}}\n\\
\lan\, f\,
\ran_{({\mathrm{A}},\mu)} \defeq & \sum_{\mathrm{a} \in {\mathrm{A}}} f(\mathrm{a})\mu(\mathrm{a})
\qquad \mathrm{when } \qquad\mu({\mathrm{A}}) = 1\,. \n
\ea
We refer to $\lan\lan\, f\,
\ran\ran_{({\mathrm{A}},\mu)}$ as the `expected sum value' of $f$, and call $\lan\, f\,
\ran_{({\mathrm{A}},\mu)}$ the `mean (average) value' of  $f$.

To emphasize distinction between spaces with measures normalized to
1 and those with not normalized measures, we use the following
convention: uppercase blackboard bold Latin letters  denote sample
spaces and the corresponding probability measures are denoted by
lowercase Greek letters,  e.g. $({\mathbb{A}},\A)$. For the spaces
with measures not normalized to 1 we use the opposite convention and
we write, e.g.~$(\Gamma,g)$.

\subsection{Some Useful Formulas }\label{A_A2}

It is well known (see 0.231.2 in \cite{gr}) that for any $ a, b \in
\mathbb{R}$ and any $ q
\in ] -1, 1[\;$  holds
\ba
\sum_{k = 0}^{\infty} (a + bk)q^{k} = \frac{a}{1 - q} +
\frac{b q}{(1 - q)^{2}}. \n
\ea
Combining that with the Newton's multinomial  formula it is easy to
obtain the following relation: for any $ a, b \in
\mathbb{R}$,  $\,  J
\in (\mathbb{N}\setminus \{0\})$, $\, F_{1},...,F_{J} \in\, ]0,1[\,$ and
$\, \sum_{j=1}^{J}F_{j} <1$
\ba
\fl
\sum_{\zeta(1)=0}^{\infty}...\sum_{\zeta(J)=0}^{\infty}\frac{|\zeta|!}{\zeta!}(a + b|\zeta|)
\prod_{j=1}^{J}F_{j}^{\zeta(j)} =  \frac{a}{1 - \sum_{j = 1}^{J}F_{j}} + \frac{b
\sum_{j = 1}^{J}F_{j}}{(1 - \sum_{j = 1}^{J}F_{j})^{2}}\;.\n 
\ea
For any $ q \in \,] -1, 1[\,$ and any $M \in
(\mathbb{N}\setminus\{0\})$ holds
\ba
\sum_{\xi(1)=0}^{\infty}...\sum_{\xi(M)
=0}^{\infty}|\xi|q^{|\xi|} = \frac{M\cdot q}{(1 - q)^{(M + 1)}} =
\frac{M \cdot q}{(1 - q)}
\sum_{\xi(1)=0}^{\infty}...\sum_{\xi(M) =0}^{\infty}q^{|\xi|}\;.\label{A3}
\ea
(\ref{A3}) can be easily verified by mathematical induction.


\section{Calculations related to $ \Gamma(N,\tau)$ } \label{A_B}
In this appendix derivations of (\ref{SN}), (\ref{FN})  and
(\ref{Ss}) from section~\ref{section3} are presented.

\subsection{Calculation of ${g}_{{\scriptscriptstyle{ (N,}}\tau
{\scriptscriptstyle{)}}}(S(N,\tau))$}\label{A_B1}

To derive (\ref{SN}) we recall that a round ends successfully if and
only if between the first and the $Nth$ successful bit flip not more
then $\tau$ application of procedure $T$ were needed
\ba
\fl
\big(\gamma\in S(N,\tau) \big) \iff
\Big(\gamma(0)\in\big(\overline{1,N}\big)
\quad \mathrm{and}\quad
\gamma(\tau)=N-\sum_{l=0}^{\tau-1}\gamma(l)=\ww(\tau)\Big)
.\n
\ea
Combining this with definition (\ref{measure_g}), we obtain
\ba
\fl \n {g}_{{\scriptscriptstyle{ (N,}}\tau {\scriptscriptstyle{)}}}(S(N,\tau))=
\sum_{\w(0)=1}^{\ww(0)} { \ww(0)\choose\w(0) } p^{\w(0)} q^{\bar\w(0)}
  \left[ \prod_{l=1}^{\tau-1}\sum_{\w(l)=0}^{\ww(l)} {\ww(l) \choose\w(l) }
   p^{\w(l)} q^{\bar\w(l)} \right]  p^{\ww(\tau)}
   \\ \n
= \sum_{\w(0)=1}^{\ww(0)} { {\ww(0)}\choose\w(0) } p^{\w(0)}
q^{\bar\w(0)}
  \left[ \prod_{l=1}^{\tau-2}\sum_{\w(l)=0}^{\ww(l)} {\ww(l) \choose\w(l) }
  p^{\w(l)} q^{\bar\w(l)} \right]\times\n\\
\times
\sum_{\w(\tau-1)=0}^{\ww(\tau-1)} {\ww(\tau-1) \choose\w(\tau-1) }
 p^{\w(\tau-1)} \big(pq\big)^{\bar\w(\tau-1)}
 \n \\
 =\sum_{\w(0)=1}^{\ww(0)} { {\ww(0)}\choose\w(0) } p^{\w(0)} q^{\bar\w(0)}
  \left[ \prod_{l=1}^{\tau-2}\sum_{\w(l)=0}^{\ww(l)} {\ww(l) \choose\w(l) }
  p^{\w(l)} q^{\bar\w(l)} \right]\Bigg(p\sum_{j=0}^{1}q^{j}\Bigg)^{\ww(\tau-1)}
  \n\\
  = \sum_{\w(0)=1}^{\ww(0)} { {\ww(0)}\choose\w(0) } p^{\w(0)} q^{\bar\w(0)}
  \left[ \prod_{l=1}^{\tau-3}\sum_{\w(l)=0}^{\ww(l)} {\ww(l) \choose\w(l) }
   p^{\w(l)} q^{\bar\w(l)} \right]\times\n \\
   \n \times
 \sum_{\w(\tau-2)=0}^{\ww(\tau-2)} {\ww(\tau-2) \choose\w(\tau-2) }
 p^{\w(\tau-2)}\Big(pq\sum_{j=0}^{1}q^{j}\Big)^{\bar\w(\tau-2)}
 \\ \n=
 \sum_{\w(0)=1}^{\ww(0)} { {\ww(0)}\choose\w(0) } p^{\w(0)} q^{\bar\w(0)}
  \left[ \prod_{l=1}^{\tau-3}\sum_{\w(l)=0}^{\ww(l)} {\ww(l) \choose\w(l) }
  p^{\w(l)} q^{\bar\w(l)} \right]
 \Bigg(p\sum_{j=0}^{2}q^{j}\Bigg)^{\ww(\tau-2)}= \\ \n = ...\,\,
 {\mathrm{ after}}\,\,{\mathrm{ the}}\,\, {\mathrm{next}} \,\, (\tau - 3)
 \,\,{\mathrm{steps}} \,\,... = \\ \n
 =\sum_{\w(0)=1}^{\ww(0)} { {\ww(0)}\choose\w(0) } p^{\w(0)} q^{\bar\w(0)}
\Bigg(p\sum_{j=0}^{\tau - 1}q^{j}\Bigg)^{\ww(1)} =
p^N\Bigg[\Big(\sum_{j=0}^\tau q^j\Big)^N-
\Big(\sum_{j=1}^\tau q^j\Big)^N \Bigg]  \\  \n =
\frac{ p^N}{(1-q)^{ N }}\Big[ (1-q^{\tau+1})^N-q^N(1-q^\tau)^N\Big]=
(1-q^{\tau+1})^N-q^N(1-q^\tau)^N\,.
\ea
We have shown that $ {g}_{{\scriptscriptstyle{ (N,}}\tau 
{\scriptscriptstyle{)}}}(S(N,\tau))=
(1-q^{\tau+1})^N-q^N(1-q^\tau)^N\,.\n
$ Comparing formula above  with   (\ref{sp}) we see that  $ 
{g}_{{\scriptscriptstyle{ (N,}}\tau
{\scriptscriptstyle{)}}}(S(N,\tau))= sp(N,\tau), \n 
$ 
which ends derivation of (\ref{SN}).

\subsection{Calculation of ${g}_{{\scriptscriptstyle{ (N,}}\tau
{\scriptscriptstyle{)}}}(F(N,\tau))$}\label{A_B2}

To derive (\ref{FN}), we recall that according to definition a round
ends in failure if and only if $\tau$ application of procedure $T$
after the first successful bit flip still at least one bit remains
in state $|0\ran$.
\ba
\fl
\big(\gamma\in F(N,\tau) \big) \iff
\Big(\gamma(0)\in\big(\overline{1,N-1}\big)
\quad \mathrm{and}\quad \forall l\in \big(\overline{1,\tau}\big)\;\;
\gamma(l)\in \big(\overline{0,\ww(l)-1}\big)  \Big)
.\n
\ea
Inserting that into (\ref{measure_g}),  we  calculate
${g}_{{\scriptscriptstyle{ (N,}}\tau
{\scriptscriptstyle{)}}}(F(N,\tau))$ and obtain, successively,
\ba
{g}_{{\scriptscriptstyle{ (N,}}\tau
{\scriptscriptstyle{)}}}(F(N,\tau))=
\sum_{\w(0)=1}^{\ww(0)-1}  {\ww(0) \choose \w(0)}  p^{\w(0)} q^{\bar\w(0)}
\prod_{l=1}^\tau
\sum_{\w(l)=0}^{{\ww(l)-1}} {{\ww(l)} \choose \w(l)} p^{\w(l) }
q^{\bar\w(l)}\n\\ \n=
\sum_{\w(0)=1}^{\ww(0)-1}  {\ww(0) \choose \w(0)}  p^{\w(0)}
q^{\bar\w(0)}\prod_{l=1}^{\tau-1}
\sum_{\w(l)=0}^{{\ww(l)-1}} {\ww(l) \choose \w(l)} p^{\w(l) }
q^{\bar\w(l)}((q+p)^{\ww(\tau)}-p^{\ww(\tau)})\n\\=
\sum_{\w(0)=1}^{\ww(0)-1}  {\ww(0) \choose \w(0)}  p^{\w(0)}
q^{\bar\w(0)}\prod_{l=1}^{\tau-1}
\sum_{\w(l)=0}^{{\ww(l)-1}} {{\ww(l)} \choose \w(l)} p^{\w(l) }
q^{\bar\w(l)}\Big[1-\Big(p
\sum_{j=0}^{0}q^{j}\Big)^{\ww(\tau)}\Big]\n\\=
\sum_{\w(0)=1}^{\ww(0)-1}  {\ww(0) \choose \w(0)}  p^{\w(0)}
q^{\bar\w(0)}\prod_{l=1}^{\tau-2}
\sum_{\w(l)=0}^{{\ww(l)-1}} {{\ww(l)} \choose \w(l)} p^{\w(l) }
q^{\bar\w(l)}\times\n\\ \n
\times \sum_{\w(\tau - 1)=0}^{{\ww(\tau -1 )-1}} {{\ww(\tau - 1)}
 \choose \w(\tau -1)} p^{\w(\tau - 1) }
q^{\bar\w(\tau - 1)}\Big[1-\Big(p
\sum_{j=0}^{0}q^{j}\Big)^{\ww(\tau)}\Big]
 \n\\ \n=
\sum_{\w(0)=1}^{\ww(0)-1}  {\ww(0) \choose \w(0)}  p^{\w(0)}
q^{\bar\w(0)}\prod_{l=1}^{\tau-2}
\sum_{\w(l)=0}^{{\ww(l)-1}} {{\ww(l)} \choose \w(l)} p^{\w(l) }
q^{\bar\w(l)}\times \\
\n \times\Big[1- p^{\ww(\tau - 1)} - \Big(p + pq
\sum_{j=0}^{0}q^{j}\Big)^{\ww(\tau - 1)} + p^{\ww(\tau - 1)}\Big]  \\ \n=
 \sum_{\w(0)=1}^{\ww(0)-1}  {\ww(0) \choose \w(0)}  p^{\w(0)}
q^{\bar\w(0)}\prod_{l=1}^{\tau-2}
\sum_{\w(l)=0}^{{\ww(l)-1}} {{\ww(l)} \choose \w(l)} p^{\w(l) }
q^{\bar\w(l)} \Big[1- \Big(p\sum_{j=0}^{1}q^{j}\Big)^{\ww(\tau - 1)}
\Big] = \\
\n =... {\mathrm{ after\,\, the \,\,next}}\,
\, (\tau - 2) \,\, {\mathrm{ steps}}... \\ \n=
\sum_{\w(0)=1}^{\ww(0)-1}  {\ww(0) \choose \w(0)}  p^{\w(0)}
q^{\bar\w(0)} \Big[1- \Big(p\sum_{j=0}^{\tau - 1}q^{j}\Big)^{\ww(
1)}
\Big] \n\\ \n=
 1 - p^N - q^N - \Big[\Big( p+pq\sum_{j=0}^{\tau-1}q^j \Big)^{N} - p^{N} -
\Big(pq \sum_{j=0}^{\tau-1} q^j\Big)^N \Big]
  \\ \n =
 (1-q^{N}) - p^{N} \Big[\Big( \sum_{j=0}^{\tau}q^j \Big)^N -
\Big( \sum_{j=1}^{\tau} q^j \Big)^N \Big]
\n\\ \n
=(1-q^{N}) - \Big[\Big( 1 - q^{\tau + 1}\Big)^{N} - q^{N}\Big( 1 -
q^{\tau}\Big)^{N}
\Big]\, .
\ea
Using (\ref{sp}), we rewrite this formula in a form: $ 
{g}_{{\scriptscriptstyle{ (N,}}\tau 
{\scriptscriptstyle{)}}}(F(N,\tau))= 1-q^N-sp(N,\tau)\,, \n $ 
which shows that (\ref{FN}) holds.


\subsection{Calculation of  ${g}_{{\scriptscriptstyle{ (N,}}
\tau {\scriptscriptstyle{)}}}( \Gamma^{\sigma}_{N}(N,\tau))$ }\label{A_B3}

For derivation of (\ref{Ss}), it is convenient to separate the
following three cases:
\\
$1)$ When  $\sigma=0$, which means that a round ended successfully at
step ${k}+1$ and, accordingly, $ \;{g}_{{\scriptscriptstyle{ 
(N,}}\tau {\scriptscriptstyle{)}}}(( \Gamma^{0}_{N}(N,\tau))=p^N = 
sp(N,0) - 
sp(N, -1).$\\
 $2)$ When $\sigma=1$, then at least one elementary 
success has to appear at step $ k+1 $ and at least one at step $ k+2 
$, thus,
\ba
\fl \n
{g}_{{\scriptscriptstyle{ (N,}}\tau
{\scriptscriptstyle{)}}}(\Gamma^{1}_{N}(N,\tau) )=
\sum_{{\w(0)}=1}^{\ww(0)-1} { {\ww(0)}\choose{\w(0)} }
 p^{{\w(0)}} (p q)^{\bar\w(0)}  =
 (p+pq)^N-p^N-(p q)^N \n \\ \n  =
  p^N\Bigg[\Bigg(\sum_{j=0}^1q^j\Bigg)^N-\Bigg(\sum_{j=1}^1q^j\Bigg)^N -1\Bigg]
= sp(N,1) - sp(N,0).
\ea
$3)$ Finally, for $ \s\in\left({\overline{2,\tau}}\right) $ we
obtain, successively,
\ba
\fl \n
{g}_{{\scriptscriptstyle{ (N,}}\tau
{\scriptscriptstyle{)}}}(\Gamma^{\sigma}_{N}(N,\tau)) =
\sum_{{\w(0)}=1}^{\ww(0)-1} { {\ww(0)}\choose{\w(0)} } p^{{\w(0)}} q^{\bar\w(0)}
\left[\prod_{l=1}^{\sigma-1}\sum_{{\w(l)}=0}^{{\ww(l)}-1} { {\ww(l)} \choose{\w(l)} } p^{\w(l)}
q^{\bar\w(l)}
\right]p^{{\ww(\s)}}\n\\ \n
=
 \sum_{{\w(0)}=1}^{\ww(0)-1} { {\ww(0)}\choose{\w(0)} } p^{{\w(0)}} q^{\bar\w(0)}
\left[\prod_{l=1}^{\sigma-2}\sum_{{\w(l)}=0}^{{\ww(l)}-1} { {\ww(l)} \choose{\w(l)} } p^{\w(l)}
q^{\bar\w(l)} \right] \times \n \\ \n
\times \Big[\Big(p\sum_{j=0}^{1}q^j\Big)^{{\ww(\s-1)}}-\Big(p\sum_{j=0}^{0}q^j\Big)^{{\ww(\s-1)}}\Big]\n
\\ \n
= \sum_{{\w(0)}=1}^{\ww(0)-1} { {\ww(0)}\choose{\w(0)} } p^{{\w(0)}}
q^{\bar\w(0)}
\left[\prod_{l=1}^{\sigma-3}\sum_{{\w(l)}=0}^{{\ww(l)}-1} { {\ww(l)} \choose{\w(l)} } p^{\w(l)}
q^{\bar\w(l)}
\right]\times\n\\ \n
\times\Big[\Big(p\sum_{j=0}^{2}q^j\Big)^{{\ww(\s-2)}}- \Big(p\sum_{j=0}^{1}q^j\Big)^{{\ww(\s-2)}}\Big]
\n \\ \n =...\; {\mathrm{ after\,\, the \,\,next}}\,
\, (\sigma - 3) \,\, {\mathrm{ steps}}\,... \\ \n
=
\sum_{{\w(0)}=1}^{\ww(0)-1} { {\ww(0)}\choose{\w(0)} } p^{{\w(0)}} q^{\bar\w(0)}
\Bigg[ \Big(p\sum_{j=0}^{\sigma-1}q^j\Big)^{{\ww(1)}}-
\Big(p\sum_{j=0}^{\sigma-2}q^j\Big)^{{\ww(1)}}
\Bigg]\n\\ \n
= p^N \left[  \Big(\sum_{j=0}^{\sigma} q^j\Big)^N-
\Big(\sum_{l=1}^{\sigma} q^j\Big)^N \right] - p^N\left[
\Big(\sum_{j=0}^{\sigma-1} q^j\Big)^N-
\Big(\sum_{j=1}^{\sigma-1} q^j\Big)^N
\right]\n\\ \n
 = sp(N,\s)-sp(N,\s-1)\,.
\ea
Comparing these three cases we see that, for  $ \s
\in\left(\overline{0,\tau}\right)$,
\ba
\label{App29}
{g}_{{\scriptscriptstyle{ (N,}}\tau {\scriptscriptstyle{)}}}( \Gamma^{\sigma}_{N}(N,\tau))
=sp(N,\s)-sp(N,\s-1)\n\\
=\Big[(1-q^{\sigma+1})^N-(1+q^N)(1-q^\sigma)^N+q^N(1-q^{\sigma-1})^N
\Big].
\ea
It is also easy to check that:
\ba
\label{App29a}
\sum_{\sigma=0}^\tau {g}_{{\scriptscriptstyle{ (N,}}\tau {\scriptscriptstyle{)}}}( \Gamma^{\sigma}_{N}(N,\tau))
=\sum_{\sigma=0}^\tau (sp(N,\s)-sp(N,\s-1))\n\\
=sp(N,\tau)-sp(N,-1)=sp(N,\tau)=g_{{\scriptscriptstyle{ (N,}}\tau
{\scriptscriptstyle{)}}}\big(S( N,\tau)\big),
\ea
and more generally, for $\tau'<\tau$,
\ba
\sum_{\sigma=0}^{\tau'} {g}_{{\scriptscriptstyle{ (N,}}\tau {\scriptscriptstyle{)}}}
( \Gamma^{\sigma}_{N}(N,\tau)) =sp(N,\tau')=g_{{\scriptscriptstyle{
(N,}}\tau' {\scriptscriptstyle{)}}}
\big(S(N,\tau')\big).
\label{appB3}
\ea
\eref{appB3} suggests that there exists a natural correspondence
between measures ${g}_{{\scriptscriptstyle{ (N,}}\tau
{\scriptscriptstyle{)}}}$ and ${g}_{{\scriptscriptstyle{ (N,}}\tau'
{\scriptscriptstyle{)}}}$  and  a simple embedding of set
$\Gamma(N,\tau')$ into $\Gamma(N,\tau)$. For details see 
{\mbox{\ref{A_B5}}}.

\subsection{Calculation of
 $\lan\lan \,\lambda\,\ran\ran_{ \left( S(N,\tau),g \right) }$  }\label{A_B4}

To derive (\ref{e14}) we combine  (\ref{App29}) and (\ref{App29a})
and obtain:
\ba
\lan\lan \,\lambda\,\ran\ran_{ \left( S(N,\tau),g \right)
}=\sum_{\sigma=0}^{\tau}\sigma {g}_{{\scriptscriptstyle{ (N,}}\tau
{\scriptscriptstyle{)}}}(
\Gamma^{\sigma}_{N}(N,\tau)) =
\sum_{j=1}^\tau \sum_{\sigma=j}^{\tau} {g}_{{\scriptscriptstyle{ (N,}}\tau {\scriptscriptstyle{)}}}(
\Gamma^{\sigma}_{N}(N,\tau))   \n \\
 =\sum_{j=1}^\tau \sum_{\sigma=j}^{\tau}(sp(N,\s)-sp(N,\s-1))=
 \sum_{j=1}^\tau (sp(N,\tau)-sp(N,j-1))
 \n \\
=\tau\cdot sp(N,\tau)-\sum_{j=0}^{\tau-1}sp(N,j). \n
\ea
It means that:
\ba\fl
\lan\lan \,\lambda\,\ran\ran_{ \left( S(N,\tau),g \right)
} =
\tau \Big[(1-q^{\tau+1})^N- q^N(1-q^\tau)^N\Big]
- \sum_{j=0}^{\tau-1}\Big[(1-q^{j+1})^N- q^N(1-q^j)^N\Big]\n\,.
\ea

\subsection{Correspondence between
measures ${g}_{{\scriptscriptstyle{ (N,}}\tau'
{\scriptscriptstyle{)}}}$ and ${g}_{{\scriptscriptstyle{ (N,}}\tau
{\scriptscriptstyle{)}}}$}\label{A_B5}

For fixed natural numbers $N, \tau,\tau'$ fulfilling conditions
$N\geq2$ and $\tau>\tau'$, there exists a natural embedding of set
$\Gamma(N,\tau')$ into $\Gamma(N,\tau)$
\ba \n
z_{\tau\tau'}:\Gamma(N,\tau')\ni \gamma \longrightarrow
z_{\tau\tau'}(\gamma)\in \Gamma(N,\tau)   \n
\ea
defined as
\ba \n
z_{\tau\tau'}(\gamma)(l)\defeq
\cases{\gamma(l)\qquad{\mathrm{for}}\qquad l\in \big(\overline{0,\tau'}\big)\\
       0 \qquad\;\;\;\;{\mathrm{for}}\qquad l\in \big(\overline{\tau'+1,\tau}\big)\,.
}
\ea
From this definition follows that, for $1\leq {\mathrm{M}}\leq N$
and $0\leq \s \leq \tau'$,
\ba \n
 z_{\tau\tau'}\big(\Gamma_{\mathrm{M}}^\s(N,\tau') \big) = \Gamma_{\mathrm{M}}^\s(N,\tau)\,.
\ea
Thus, we can write the following relation between measures
${g}_{{\scriptscriptstyle{ (N,}}\tau' {\scriptscriptstyle{)}}}$  and
${g}_{{\scriptscriptstyle{ (N,}}\tau {\scriptscriptstyle{)}}}$
\ba \n
{g}_{{\scriptscriptstyle{ (N,}}\tau {\scriptscriptstyle{)}}}
\big(\Gamma_{\mathrm{M}}^\s(N,\tau)  \big)=
{g}_{{\scriptscriptstyle{ (N,}}\tau {\scriptscriptstyle{)}}}
\big( z_{\tau\tau'}\big(\Gamma_{\mathrm{M}}^\s(N,\tau') \big) \big)=
q^{(N-{\mathrm{M}})(\tau-\tau')} {g}_{{\scriptscriptstyle{
(N,}}\tau' {\scriptscriptstyle{)}}}
\big(\Gamma_{\mathrm{M}}^\s(N,\tau')  \big),
\ea
which is valid for $0\leq \s \leq \tau'$.  In particular,  for
$\mathrm{M}=N$,   identity
\ba
{g}_{{\scriptscriptstyle{ (N,}}\tau {\scriptscriptstyle{)}}}
\big(\Gamma_N^\s(N,\tau)  \big)=
{g}_{{\scriptscriptstyle{ (N,}}\tau' {\scriptscriptstyle{)}}}
\big(\Gamma_N^\s(N,\tau')  \big)\n\,
\ea
is obtained. From that follows:
\ba \n
{g}_{{\scriptscriptstyle{ (N,}}\tau {\scriptscriptstyle{)}}}
\Big( \bigcup_{\s=0}^{\tau'} \Gamma_N^\s(N,\tau)  \Big)=
{g}_{{\scriptscriptstyle{ (N,}}\tau' {\scriptscriptstyle{)}}}
\big(S(N,\tau')  \big)=sp(N,\tau'),
\ea
which gives exactly (\ref{appB3}).


\section{Calculations related to $({\mathbb{D}}(N,\tau),\delta) $}\label{A_C}

\subsection{Additional conventions used for a compact notation}\label{A_C1}

Note, that  set $\big(\overline{0,|\z|} \big)$ might be rewritten as
the following sum of disjoint sets:
\ba\n
\{ 0 \} \cup \big(\overline{1,\z(1)} \big) \cup \big(\overline{\z(1)+1,\z(2)} \big)\cup
\dots \cup  \big(\overline{\z(J-1)+1,\z(J)} \big)\,.
\ea
For every map {\mbox{$\xi\in
{\mathrm{Map}}\big(\big(\overline{0,|\z|}\big),{\mathbb{N}}\big)$}}
we define:
\ba\n
\fl
\xi_0\defeq\xi\bigg|_{\{0 \}}\qquad
\underline{\xi}_1 \defeq \xi\Big|_{(\overline{1,\z(1)})}
\qquad
\underline{\xi}_2 \defeq \xi\Big|_{(\overline{\z(1)+1,\z(2)})}
\quad\dots\qquad
\underline{\xi}_J \defeq \xi\Big|_{(\overline{\z(J-1)+1,\z(J)})},
\ea
where  $\xi\Big|_{(\overline{a,b})} $ denotes restriction of a map
$\xi$ to the set $(\overline{a,b}) $. This notation allows to write
the following, valid for  $j\in \big(\overline{1,J-1} \big)$,
abbreviations for sums over $\xi$
\ba
\sum_{\underline{\xi}_{(j+1)}=0}^\infty \defeq \sum_{\xi\left(\z(j)+1\right)=0}^\infty
\sum_{\xi\left(\z(j)+2\right)=0}^\infty
\dots \sum_{\xi\left(\z(j+1)\right)=0}^\infty \;. \n \quad
\ea
Accordingly,
\ba
\fl \sum_{\underline{\xi}_{(j+1)}=0}^\infty q^{N |\underline{\xi}_{(j+1)}| } \defeq
 \sum_{\xi\left(\z(j)+1\right)=0}^\infty q^{N \xi \left( \z(j)+ 1 \right) }
 \sum_{\xi\left(\z(j)+2\right)=0}^\infty q^{N \xi \left(\z(j)+2 \right)}
\dots \sum_{\xi\left(\z(j+1)\right)=0}^\infty q^{N \xi \left(\z(j+1)\right) }\;.
 \n
\ea
For $j=0$, the corresponding formulas  are given by
\ba\n
\sum_{\underline{\xi}_1=0}^\infty \defeq \sum_{\xi(1)=0}^\infty \sum_{\xi(2)=0}^\infty \dots \sum_{\xi\left(\z(1)\right)=0}^\infty
\ea
and
\ba\n  \sum_{\underline{\xi}_1=0}^\infty
q^{N |\underline{\xi}_1|} \defeq
\sum_{\xi(1)=0}^\infty q^{N\xi(1)} \sum_{\xi(2)=0}^\infty q^{N\xi(2)} \dots \sum_{\xi\left(\z(1)\right)=0}^\infty
q^{N \xi\left(\z(1)\right)} \;.
\ea
These definitions are introduced to help presenting calculations  in
a relatively compact form. For the same reason we define $
\vec{\xi{\,}}\defeq \xi\Big|_{(\overline{1,\z(J)})}\,
$ and  note that $|\xi|$ can be expressed as
\ba\n
|\xi|
=\xi(0)+|\vec{\xi{\,}}|=\xi(0)+|\underline{\xi}_1|+|\underline{\xi}_2|+\dots
+ |\underline{\xi}_J|\,.
\ea


\subsection{Probabilities $\delta\big( {\mathbb{D}}(N,\tau) \big)$}\label{A_C2}

To calculate (\ref{nD}), we use formulas and conventions from
\ref{A_A} and the notation introduced in
\ref{A_C1}. We obtain:
\ba
\fl\n
\delta\big( {\mathbb{D}}(N,\tau) \big)= \sum_{(\xi,\z,\gamma)\in{\mathbb{D}}(N,\tau) }
\delta(\xi,\zeta,\gamma)=\\
\fl \n
=\sum_{\z(1)=0}^\infty \dots \sum_{\z(J)=0}^\infty
 \sum_{\xi(0)=0}^\infty\sum_{\gamma\in S(N,\tau)}\sum_{\underline{\xi}_1=0}^\infty
\sum_{\underline{\xi}_2=0}^\infty\dots \sum_{\underline{\xi}_J=0}^\infty
  g(\gamma)  q^{N \xi(0)}
 q^{N |\underline{\xi}_1|}   \dots
q^{N |\underline{\xi}_J|}\times\n\\
\times  \frac{|\z|!}{\z!} g(f_1)^{\zeta(1)}\dots
g(f_J)^{\zeta(J)} \n\\
\fl
=\sum_{\z(1)=0}^\infty \dots \sum_{\z(J)=0}^\infty \frac{|\z|!}{\z!}
\Big( \sum_{\xi(0)=0}^\infty q^{N \xi(0)} \sum_{\gamma\in S(N,\tau)} g(\gamma)\Big)
\Big(\sum_{\underline{\xi}_1=0}^\infty q^{N |\underline{\xi}_1|} g(f_1)^{\zeta(1)}\Big)\times\n\\
\times
\Big(\sum_{\underline{\xi}_2=0}^\infty q^{N |\underline{\xi}_2|} g(f_2)^{\zeta(2)}\Big)
\dots
\Big(\sum_{\underline{\xi}_J=0}^\infty q^{N |\underline{\xi}_J|}
g(f_J)^{\zeta(J)} \Big) \n\\ \n
\fl
=\B\big({\mathbb{S_B}}(N,\tau) \big) \sum_{\z(1)=0}^\infty \dots
\sum_{\z(J)=0}^\infty \frac{|\z|!}{\z!}
 \,\beta({\cal{F}}_1)^{\z(1)}\dots{\beta(\cal{F}}_J)^{\z(J)} \\ \n
={\cal{P}}(N, \tau) \Big(1-\sum_{j=0}^J{\beta(\cal{F}}_j)\Big)^{-1}=
{\cal{P}}(N ,\tau) \big(1-{\cal{Q}(N,\tau)} \big)^{-1}=1\n,
\ea
where ${\cal{P}}(N ,\tau)$ and ${\cal{Q}}(N ,\tau)$ were defined by
(\ref{defP}) and (\ref{defQ}), respectively.


\subsection{Average values $\lan\, {\mathcal{K}}\, \ran_
{({\mathbb{D}}(N,\tau), \delta)}$}\label{A_C3}

To start derivation of (\ref{Kavr}), let us note that from
({\ref{Kdef}) follows identity:
\ba
\fl
\lan\, {\mathcal{K}}\, \ran_{({\mathbb{D}}(N,\tau),\delta)} \defeq
\sum_{(\xi,\z,\gamma)\in{\mathbb{D}}(N,\tau) }  {\cal{K}}(\xi,\zeta,\gamma)
\delta(\xi,\zeta,\gamma)\n\\
\fl
=\!\!\!\! \sum_{(\xi,\z,\gamma)\in{\mathbb{D}}(N,\tau) }
\!\!\!\!\!\!\!\!  |\vec{\xi{\,}}|
\delta(\xi,\zeta,\gamma)
+
(\tau+1)\!\!\!\!\!\!\!\!\sum_{(\xi,\z,\gamma)\in{\mathbb{D}}(N,\tau)
} \!\!\!\!\!\!\!\!\! |\zeta|
\delta(\xi,\zeta,\gamma)\;
+\!\!\!\!\!\!\!\!\sum_{(\xi,\z,\gamma)\in{\mathbb{D}}(N,\tau) }
\!\!\!\!\!\!\!\!(\xi(0)+1+\lambda(\gamma))
\delta(\xi,\zeta,\gamma). \n
\ea
For clarity, we calculate each of these sums   separately. As before,
we use formulas and conventions from
\ref{A_A} and notation introduced in
\ref{A_C1}.
The first sum is equal to
\ba \fl
\sum_{(\xi,\z,\gamma)\in{\mathbb{D}}(N,\tau) } \!\!\!\!\!\!  |\vec{\xi{\,}}|
\delta(\xi,\zeta,\gamma)=\n\\
\fl
=\sum_{\z(1)=0}^\infty \dots \sum_{\z(J)=0}^\infty \frac{|\z|!}{\z!}
\Big( \sum_{\xi(0)=0}^\infty q^{N \xi(0)} \sum_{\gamma\in S(N,\tau)} g(\gamma)\Big)
\sum_{\underline{\xi}_1=0}^\infty q^{N |\underline{\xi}_1|}\dots
\sum_{\underline{\xi}_J=0}^\infty q^{N |\underline{\xi}_J|}  |\vec{\xi{\,}}|
\prod_{j=1}^J g(f_j)^{\zeta(j)} \n\\
={\cal{P}}(N,\tau) \sum_{\z(1)=0}^\infty \dots \sum_{\z(J)=0}^\infty
\frac{|\z|!}{\z!} \prod_{j=1}^J g(f_j)^{\z(j)}
\Big[
\sum_{\underline{\xi}_1=0}^\infty q^{N |\underline{\xi}_1|}\dots
\sum_{\underline{\xi}_J=0}^\infty q^{N |\underline{\xi}_J|}  |\vec{\xi{\,}}|
\Big]\n\\
={\cal{P}}(N,\tau)\sum_{\z(1)=0}^\infty\dots\sum_{\z(J)=0}^\infty\frac{|\z|!}{\z!}
\Big(\prod_{j=1}^J g(f_j)^{\zeta(j)}\Big)
|\z|q^N \big(1-q^N\big)^{-(|\z|+1)}\n\\
={\cal{P}}(N,\tau)q^N \big(1-q^N\big)^{-1}
\sum_{\z(1)=0}^\infty\dots\sum_{\z(J)=0}^\infty\frac{|\z|!}{\z!} |\z|\,
\beta({\cal{F}}_1)^{\z(1)}\dots\beta({\cal{F}}_J)^{\z(J)}\n\\
\n = q^N
\big(1-q^N\big)^{-1} {\cal{Q}}(N,\tau) {\cal{P}}(N,\tau)^{-1}.
\ea
The second sum is given by
\ba \fl
(\tau+1)\!\!\!\!\!\!\!\!\sum_{(\xi,\z,\gamma)\in{\mathbb{D}}(N,\tau)
} \!\!\!\!\!\! |\zeta|
\delta(\xi,\zeta,\gamma)=(\tau+1)\sum_{\z(1)=0}^\infty\dots\sum_{\z(J)=0}^\infty \frac{|\z|!}{\z!} |\z|\times\n\\
\times
\Big(\sum_{\xi(0)=0}^\infty q^{N \xi(0)}\sum_{\gamma\in S(N,\tau)}g(\gamma)\Big)
\sum_{\underline{\xi}_1=0}^\infty q^{N |\underline{\xi}_1|}g(f_1)^{\zeta(1)}\dots \sum_{\underline{\xi}_J=0}^\infty q^{N |\underline{\xi}_J|}
 g(f_J)^{\zeta(J)} \n\\
=(\tau+1){\cal{P}}(N,\tau)\sum_{\z(1)=0}^\infty\dots\sum_{\z(J)=0}^\infty
\frac{|\z|!}{\z!} |\z|\,
\beta({\cal{F}}_1)^{\z(1)}\dots\beta({\cal{F}}_J)^{\z(J)}\n\\ 
=(\tau+1){\cal{Q}}(N,\tau) {\cal{P}}(N,\tau)^{-1}. \label{component2}
\ea
The third sum can be calculated as
\ba
\fl
\sum_{(\xi,\z,\gamma)\in{\mathbb{D}}(N,\tau) } \!\!\!\!\!\!(\xi(0)+1+\lambda(\gamma))
\delta(\xi,\zeta,\gamma)
=\sum_{\z(1)=0}^\infty\dots\sum_{\z(J)=0}^\infty \frac{|\z|!}{\z!}
\times \n \\
\fl
\times\Big(\sum_{\xi(0)=0}^\infty q^{N \xi(0)} \!\!\!\!\!
 \sum_{\gamma\in S(N,\tau)}g(\gamma)
\big[\xi(0)+1+\lambda(\gamma)\big]\Big)
\sum_{\underline{\xi}_1=0}^\infty q^{N |\underline{\xi}_1|}\dots
\sum_{\underline{\xi}_J=0}^\infty q^{N |\underline{\xi}_J|}
\prod_{j=1}^J g(f_j)^{\zeta(j)}\n\\ \n
= \lan\lan\, \Lambda\, \ran\ran_{({\mathbb{S_B}}(N,\tau), \beta)}
\sum_{\z(1)=0}^\infty\dots\sum_{\z(J)=0}^\infty \frac{|\z|!}{\z!} \,
\beta({\cal{F}}_1)^{\z(1)}\dots\beta({\cal{F}}_J)^{\z(J)} \\ \n
=\lan\lan\, \Lambda\,
\ran\ran_{({\mathbb{S_B}}(N,\tau), \beta)} {\cal{P}}(N,\tau)^{-1}.\n
\ea
Finally, combining these three expressions we obtain
\ba \fl
\lan\, {\mathcal{K}}\, \ran_{({\mathbb{D}}(N,\tau),\delta)} =
\lan\lan\, \Lambda\,\ran \ran_{({\mathbb{S_B}}(N,\tau),\beta)} {\cal{P}}
(N,\tau)^{-1} +
\Big(\tau+\big(1-q^N\big)^{-1} \Big) {\cal{Q}(N,\tau)} {\cal{P}}(N,\tau)^{-1}
\n\\ \n
=\lan\lan\, \Lambda\, \ran\ran_{({\mathbb{S_B}}(N,\tau), \beta)}
{\cal{P}}(N,\tau)^{-1}+
\lan\lan\, \Lambda\, \ran\ran_{({\mathbb{F_B}}(N,\tau), \beta)} {\cal{P}}(N,\tau)^{-1} \\ \n
= \lan\, \Lambda\, \ran_{({\mathbb{B}}(N,\tau), \beta)}
{\cal{P}}(N,\tau)^{-1}.
\ea
Explicit formulas for  $\lan\, \Lambda\,
\ran_{({\mathbb{B}}(N,\tau), \beta)}$ and ${\cal{P}}(N,\tau)$
can be found in section \ref{section4}.


\section{The Limit
$\displaystyle\lim_{\tau\rightarrow \infty}
\lan\, {\mathcal{K}}\, \ran_{({\mathbb{D}}(N,\tau),\delta)}$ for $N \geq 2$}\label{AppLimit}

To derive (\ref{limit}), it is convenient to treat nominator and
denominator of  $\lan\, {\mathcal{K}}\,
\ran_{({\mathbb{D}}(N,\tau),\delta)}$ separately. A denominator of
$\lan\, {\mathcal{K}}\,
\ran_{({\mathbb{D}}(N,\tau),\delta)}$ equals to  $sp(N,\tau)$;
it can be rewritten as
\ba \n \fl
sp(N,\tau) &=&(1-q^{\tau+1})^N-q^N(1-q^\tau)^N \n\\
&=&\sum_{n=0}^N (-1)^n {N \choose n} q^{(\tau+1)n} - q^N \sum_{n=0}^N (-1)^n {N \choose n} q^{\tau n} \n\\
\fl
&=&\sum_{n=0}^N (-1)^n {N \choose n} q^{\tau n} (q^n-q^N)\n\\
\n &=&
1-q^N+ \sum_{n=1}^N (-1)^n {N \choose n} q^{\tau n} \big(q^n-q^N
\big).
\ea
Because $1>q>0$,
\ba
\lim_{\tau\rightarrow\infty} {sp(N,\tau)}=1-q^N\,. \label{AppE3}
\ea
Let  $L(N,\tau)$ denotes a nominator of $\lan\, {\mathcal{K}}\,
\ran_{({\mathbb{D}}(N,\tau),\delta)}$, i.e.,
\ba
\n
L(N,\tau)\defeq 1-(1-q^\tau)^N +
(1-q^N)\Big[\tau-\sum_{j=1}^{\tau-1}(1-q^j)^N \Big]\,.
\ea
To calculate $ \displaystyle
\lim_{\tau\rightarrow\infty} L(N,\tau)$,  we rewrite formula above
using  relations
\ba \n
1-(1-q^\tau)^N=\sum_{n=1}^N (-1)^{n+1}{N \choose n} q^{\tau n}
\ea
and
\ba
\fl
\tau-\sum_{j=1}^{\tau-1}(1-q^j)^N=\tau -\sum_{j=1}^{\tau-1}\sum_{n=0}^N (-1)^n {N \choose n} q^{j n}
=\tau -\sum_{j=1}^{\tau-1} \Big[ 1+ \sum_{n=1}^N (-1)^n {N \choose n} q^{j n} \Big] \n\\
\fl \n
=1- \sum_{n=1}^N (-1)^n {N \choose n} \sum_{j=1}^{\tau-1} q^{j n} =
1- \;\sum_{n=1}^N (-1)^{n} {N \choose n} q^n \frac{1-q^{\tau
n}}{1-q^n}\,,
\ea
valid for  $\tau\geq 2$.  (When calculating limit
${\tau\rightarrow\infty}$, this condition  is fulfilled.) We obtain:
\ba
\fl \n
L(N,\tau)=\sum_{n=1}^N (-1)^{n+1}{N \choose n} q^{\tau n}+
(1-q^N)\left[ 1- \;\sum_{n=1}^N (-1)^{n} {N \choose n} q^n
\frac{1-q^{\tau n}}{1-q^n} \right]
\ea
and the limit of $L(N,\tau)$ is given by
\ba
\lim_{\tau\rightarrow\infty}  L(N,\tau)
= (1-q^N) \sum_{n=1}^N (-1)^{(n + 1)} {N \choose n}
\frac{1}{1-q^n}\label{AppE5}.
\ea
From (\ref{AppE3}) and (\ref{AppE5}) follows that:
\ba \n
\lim_{\tau\rightarrow\infty}  \lan\, {\mathcal{K}}\, \ran_{({\mathbb{D}}(N,\tau), \delta)}=
\sum_{n=1}^N (-1)^{(n + 1)}{N \choose n}  \frac{1}{1-q^n}\;.
\ea

\paragraph{Remark 4:}

Calculations presented above allow to write (\ref{nowe3}) in a 
different form:
\ba
\fl
\lan\, {\mathcal{K}}\, \ran_{({\mathbb{D}}(N,\tau), \delta)}=
\frac{(1-q^N)- \sum_{n=1}^N (-1)^{n } {N \choose n} \Big[q^{\tau n}
+ q^n (1-q^N) (1-q^{\tau n})(1-q^n)^{-1} \Big]} {1-q^N
+\sum_{n=1}^{N-1} (-1)^{n} {N \choose n} q^{\tau n}
(q^n-q^N)}\label{inne3}.
\ea
Right-hand-side of this formula is well defined for $\tau
\in
\mathbb{R}$, but we should remember that
$\lan\, {\mathcal{K}}\, \ran_{({\mathbb{D}}(N,\tau), \delta)} $ was
derived for $N,\tau\in(\mathbb{N}\setminus \{0,1\})$.
$\quad{\scriptscriptstyle{\square}}$

\section{Toy model}  \label{A_E}
In order to associate a more intuitive meaning to (\ref{Kavr}), let
us consider the following: for any two real numbers $0 < P < 1$,
$Q=1-P$ and a map $\varrho_{\scriptscriptstyle{P}}$ defined by
\ba
\varrho_{\scriptscriptstyle{P}}:{\mathbb{N}}\ni n
\longrightarrow \varrho_{\scriptscriptstyle{P}}(n)
\defeq {Q}^N {P} \in[0,1] , \n
\ea
pair $(\mathbb{N},\varrho_{\scriptscriptstyle{P}})$  defines a
probability space. Introducing also two real parameters
$\mathcal{A}$, $
\mathcal{B}$ and a random variable
 $\mathcal{L}_{\mathcal{A}\mathcal{B}}:{\mathbb{N}}\ni n \rightarrow
{\cal{L}}_{\mathcal{A}\mathcal{B}} (n)\defeq {\mathcal{A}}+n
{\mathcal{B} } \in {\mathbb{R}} $, we can easily check that:
\ba
\n
\lan {\cal{L}}_{\mathcal{A}\mathcal{B}} 
\ran_{(\mathbb{N},{\varrho_P})}=
{\cal{A}}+{\cal{B }} Q/P
\ea
For $P={\cal{P}}(N,\tau)$, $Q={\cal{Q}}(N,\tau)$ defined by 
(\ref{defP}) and (\ref{defQ}) and  ${\cal{A}}$, ${\cal{B}}$ 
{\mbox{equal to:}}
\ba
{\cal{A}}= \frac{\lan\lan\,\Lambda
\,\ran\ran_{(\mathbb{S_B}(N,\tau),\B)}}{{\cal{P}}(N,\tau)}\qquad {\mathrm{and}}\qquad
{\cal{B}}=\frac{\lan\lan\,\Lambda
\,\ran\ran_{(\mathbb{F_N}(N,\tau),\B)}}{{\cal{Q}}(N,\tau) } ,\n
\ea
we obtain:
\ba
\n
\lan {\cal{L}}_{\mathcal{A}\mathcal{B}} \ran_{(\mathbb{N},\varrho_P)}=
\frac{\lan\lan\,\Lambda
\,\ran\ran_{(\mathbb{S_B}(N,\tau),\B)}}{{\cal{P}}(N,\tau) }+
\frac{\lan\lan\,\Lambda
\,\ran\ran_{(\mathbb{F_B}(N,\tau),\B)}}{{\cal{P}}(N,\tau) }\\=
\frac{\lan\,\Lambda
\,\ran_{(\mathbb{B}(N,\tau),\B)}}{{\cal{P}}(N,\tau) }
=\lan\, {\mathcal{K}}\, \ran_{({\mathbb{D}}(N,\tau),\delta)}\,.\n
\ea
${\cal{A}}$ denotes an average number of steps in a round that ended
successfully, ${\cal{B}}$ is an average number of steps in a round
that ended unsuccessfully, $Q/P$ is an average number of rounds that
ended unsuccessfully for a round that ended successfully:

\noindent
Average number of steps = (average number of steps in a round that
ended successfully) + (average number of steps in a round that ended
unsuccessfully)$\times$ (average number of rounds that ended
unsuccessfully for a round that ended successfully)

To include a non-zero reset time $\Omega$, we insert
$\mathcal{B}_\Omega=\mathcal{B}+\Omega$ instead of $\mathcal{B}$ in
the relations above and reproduce results from section
\ref{section5_3}.


\end{document}